\documentclass[dvips]{article}

\usepackage{icrc2011}
\usepackage{color}

\title{The Microwave Air Yield Beam Experiment (MAYBE): measurement of GHz
radiation for Ultra-High Energy Cosmic Rays detection}

\newcommand{\etal}{\MakeLowercase{\textit{et al. }}} 
\shorttitle{M.~Monasor \etal MAYBE: measurement of GHz radiation for UHECR detection}

\authors{M.~Monasor$^{1}$, M.~Boh\'{a}\v{c}ov\'{a}$^{2}$, C.~Bonifazi$^{3}$, G.~Cataldi$^{4}$, S.~Chemerisov$^{5}$, J.R.T.~de Mello Neto$^{3}$, P. Facal San Luis$^{1}$, B.~Fox$^{6}$, P.W.~Gorham$^{6}$, C.~Hojvat$^{7}$, N.~Hollon$^{1}$, R.~Meyhandan$^{6}$, L.C.~Reyes$^{1}$, B.~Rouill\'{e} d'Orfeuil$^{1}$, E.M.~Santos$^{3}$, J.~Pochez$^{1}$, P.~Privitera$^{1}$, H.~Spinka$^{5,\:6}$, V.~Verzi$^{8}$, C.~Williams$^{1}$, J.~Zhou$^{1}$}
\afiliations{$^1$ Kavli Institute for Cosmological Physics, University of Chicago IL 60637, USA\\ $^2$ Institute of Physics, Academy of Sciences of the Czech Republic, Prague, Czech Republic\\$^3$ Universidade Federal do Rio de Janeiro, Instituto de F\'{\i}sica, Rio de Janeiro, Brazil\\$^4$ Dipartimento di Fisica dell'Universit\`{a} del Salento and Sezione INFN, Lecce, Italy \\$^5$ Argonne National Laboratory, Argonne, IL, USA \\$^{6}$ Department of Physics and Astronomy, University of Hawaii at Manoa, Honolulu, HI, USA \\$^{7}$ Fermilab, Batavia, IL, USA \\$^8$ Universit\`{a} di Roma II "Tor Vergata" and Sezione INFN, Roma, Italy}

\email{monasor@uchicago.edu}

\abstract{We present first measurements by MAYBE of microwave emission from an electron  beam induced air plasma, performed at the electron Van de Graaff facility of the Argonne National Laboratory.  Coherent radio Cherenkov, a major background in a previous beam experiment, is not produced by the 3 MeV beam, which simplifies the interpretation of the data. Radio emission is studied over a wide range of frequencies between 3 and 12 GHz. This measurement provides further insight on microwave emission from extensive air showers as a novel detection technique for Ultra-High Energy Cosmic Rays.}

\begin{document}
\maketitle

\section{Introduction}

The possibility of studying air showers induced by cosmic rays in the microwave range has motivated the development of several prototypes~\cite{microwave_auger} to study the feasibility of a new detection technique that would combine the major advantage of the fluorescence technique, i.e. observation of the shower development, with a 100\% duty cycle and almost negligible atmospheric attenuation. Recent laboratory measurements~\cite{gorham} have found evidence of coherent broadband microwave emission from a laboratory plasma created ionizing air with high-energy electrons. This emission has been interpreted as molecular bremsstrahlung radiation with a major departure from the steady-state thermal emission limit. This is due to the particular conditions recreated in the plasma, i.e specific electron density or geometrical configuration, that can induce phase-space correlations and therefore enhance the floor level signal.

Due to the difficulty in quantifying analytically the expected emission and its dependence on electron density, laboratory measurements under different plasma conditions would be desirable. We have performed an accelerator-based experiment designed to characterize the microwave emission from a plasma created using a 3 MeV electron beam injected into a RF anechoic chamber of 1 m$^3$ size. It was carried out at the Van der Graaff accelerator facility at Argonne National Laboratory. A detailed description of the experiment and first measurements are described in this contribution. 

\section{Experimental setup}


\subsection{Electron beam}

The beam consists of bunches of 3 MeV electrons with adjustable pulse width from 5 ns to 1 ms. The electron current and repetition rate can also be adjusted to generate bunches containing typically from $\sim10^{10}$ to $\sim10^{13}$ electrons. The transverse size of the beam is $\sim$5 mm and the exit window at the end of the beam pipe is a Duraluminum foil of 0.002" thickness. Notice the electron energy is well below the threshold for Cherenkov production in air, a major source of  background radiation in a previous experiment~\cite{gorham}. A pickup coil was placed at the exit window of the beam pipe to monitor the beam current. 


\subsection{Anechoic chamber and antennas}

\begin{figure}[!t]
  \centering
  \includegraphics[width=0.85\columnwidth]{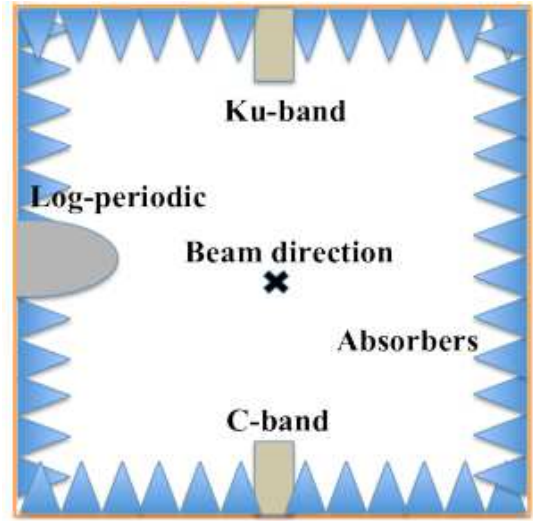}
  \caption{Schematic of anechoic chamber and layout of the radio receivers.}
  \label{layout}
 \end{figure}

A schematic of the layout is presented in figure~\ref{layout}. The beam enters a copper anechoic Faraday chamber of $\sim1~{\rm m}^3$ filled with air that prevents external radiation from getting into the inner antennas. The inner surface is covered with pyramidal absorbers that provide a minimum of 30 dB absorption at normal incidence for frequencies above 1 GHz. A circular port of 3 cm diameter  was opened in the chamber to allow the entry of the 3 MeV electrons that otherwise would not penetrate the copper wall. In this configuration, transition radiation generated at the Duraluminum window and entering the chamber through this aperture can be a source of background and must be taken into account. 

Three different radio receivers were mounted inside the chamber to characterize the microwave emission in a broad frequency spectrum. The first two were commercial C-band (3.4-4.2 GHz) and Ku-band (12.2-12.7 GHz) feeds which system temperatures and gains were previously calibrated in the lab. 
The third receiver was a Rohde \& Schwarz (R\&S) HL050 log-periodic antenna. Due to its broadband input frequency (0.85-26.5 GHz) this antenna is suitable to measure the microwave emission over a wide spectral range. At the time of this beam test the log-periodic antenna was connected to a 1-2 GHz low noise amplifier with an average gain of 40 dB so the spectral region studied with this receiver was reduced to what we denominated here low frequency range. All the antennas are located in a central plane perpendicular to the beam direction and pointing towards it so the distance beam axis-antenna is always $\sim$0.5 m. Due to the limitation in distance, the response of the receivers could be affected by near-field effects. The receivers only measure one polarization at a time. Both polarizations, co-polarized and cross-polarized with the beam axis, can be recorded. C-band, Ku-band and low frequency (0.7-2.4 GHz) receivers were also located outside the chamber and close to the beam exit to monitor external radiation. Signals from all antennas were transmitted to the control room through $\sim14$ m coaxial cables that have an average loss of 0.3 dB per meter in the frequency range of the transmitting signal. Losses from cable, adapters and connectors must be properly evaluated to later correct the recorded signal. 


\subsection{Simulations}

\begin{figure*}[!t]
  \centerline{\includegraphics[height=2.9in]{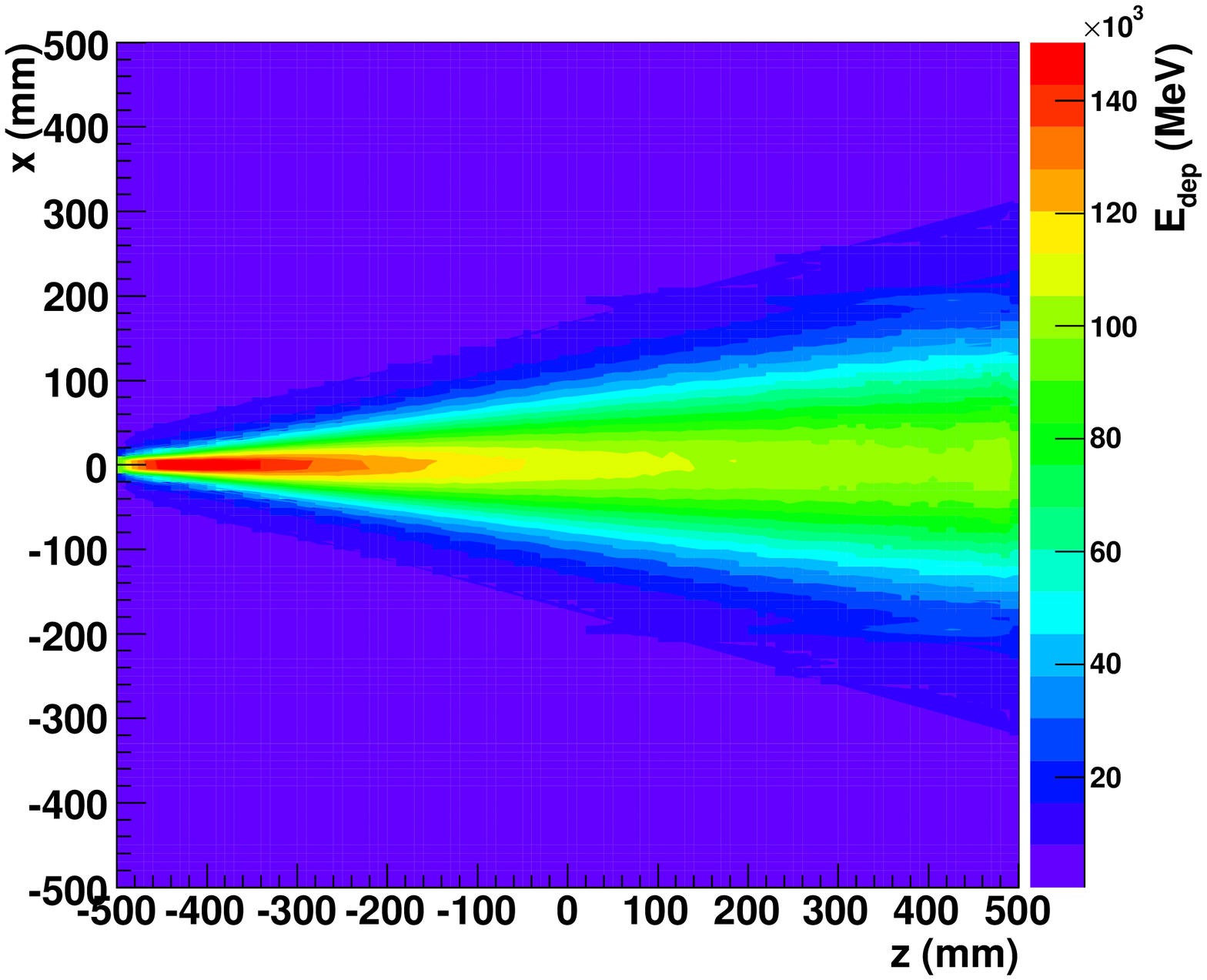}\label{fig2a}
    \hfil
    \includegraphics[height=2.9in]{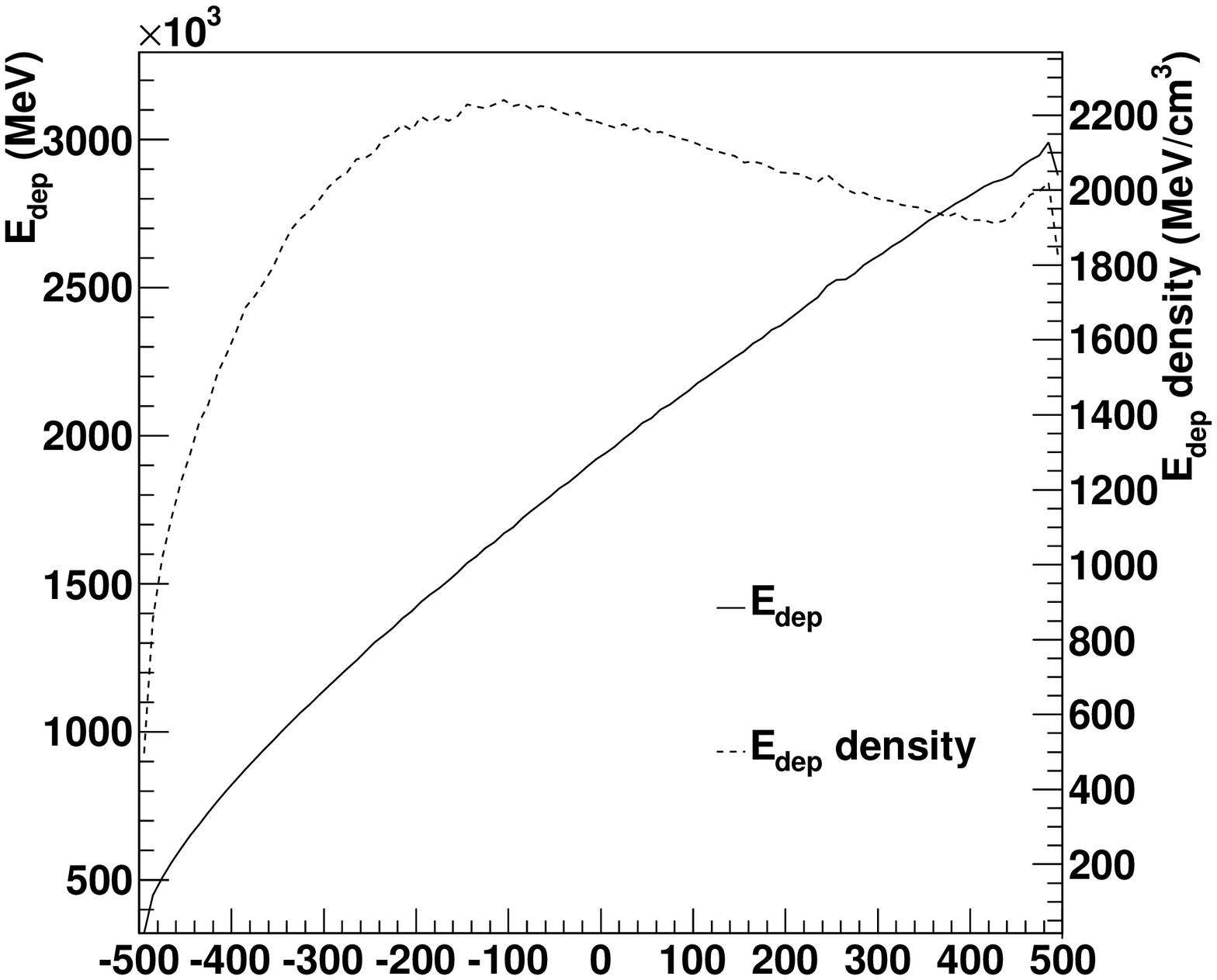} \label{fig2b}
  }
  \caption{Geant 4 simulations of the energy deposit inside the chamber. Left: Energy deposit as a function of longitudinal and transversal dimensions. Right: longitudinal development of energy deposit and its density.
  }
  \label{simulations}
\end{figure*}

Geant4 simulations reproducing the setup and running conditions were performed in order to characterize the plasma created inside the chamber. The energy deposit in the chamber allows us to get a quantitative idea of the number of ionization electrons contained in the generated plasma and its density. The left panel of figure~\ref{simulations} shows the energy deposit in the chamber as a function of both longitudinal $z$ and transversal $x$ directions (the distribution in $y$ is symmetrical to the one in $x$) for a typical beam configuration with $2.5\times10^{9}$ electrons. The bin area selected to integrate the energy deposit is 1cm x 1cm. As it can be observed,  the energy deposit is not uniformly distributed in the chamber but denser in a conical region centered in the beam trajectory with a radius increasing from a few mm to $\sim$18 cm. The right panel of figure~\ref{simulations} shows the integrated energy deposit as a function of $z$ and its density calculated in cylindrical slices of 1 cm height and radius given by $x_{\rm RMS}$. This density, that increases fast in the first portion of the chamber and then remains almost constant, will be proportional to the ionization electron density. Assuming all the energy deposit is invested in ionization we obtain typical values for the ionization electron density of $\sim10^{8}$ electrons per cm$^3$.


The total energy deposit in the chamber calculated over the number of electrons pulsed in 3ns\footnote{the measured signal comes from the contribution of electrons inside the chamber at a certain time that can be calculated from the time the electrons need to cross the 1 m long chamber.}  will oscillate for the different running conditions between $10^{14}$ and $10^{15}$ eV.  This amount of energy deposit is comparable to the total energy deposit at the depth of maximum development by an extensive air shower originated by a cosmic ray of energy $\sim10^{18}$ eV.

\section{Data acquisition}

Data were taken both in frequency and time domains using a R\&S spectrum analyzer (9 kHz-30 GHz frequency range) and a Tektronix TDS6154C oscilloscope (40 GS/s and 20 GHz bandwidth) respectively. 
Coinciding with the beginning and end of the beam entering the chamber we observed a prompt and strong peak of signal in the log-periodic antenna whose origin is still unknown. A narrower peak is also occasionally observed at the beginning of the trace in the C-band receiver, most noticeable at high beam energies. This effect could be due to some charge deposit on the chamber or external noise. Apart from this peak, a clear signal is measured in coincidence with the beam in both receivers. 
The analysis for the Ku-band antenna is still ongoing.


\section{Data analysis}

\begin{figure*}[!t]
  \centerline{\includegraphics[height=0.7\columnwidth]{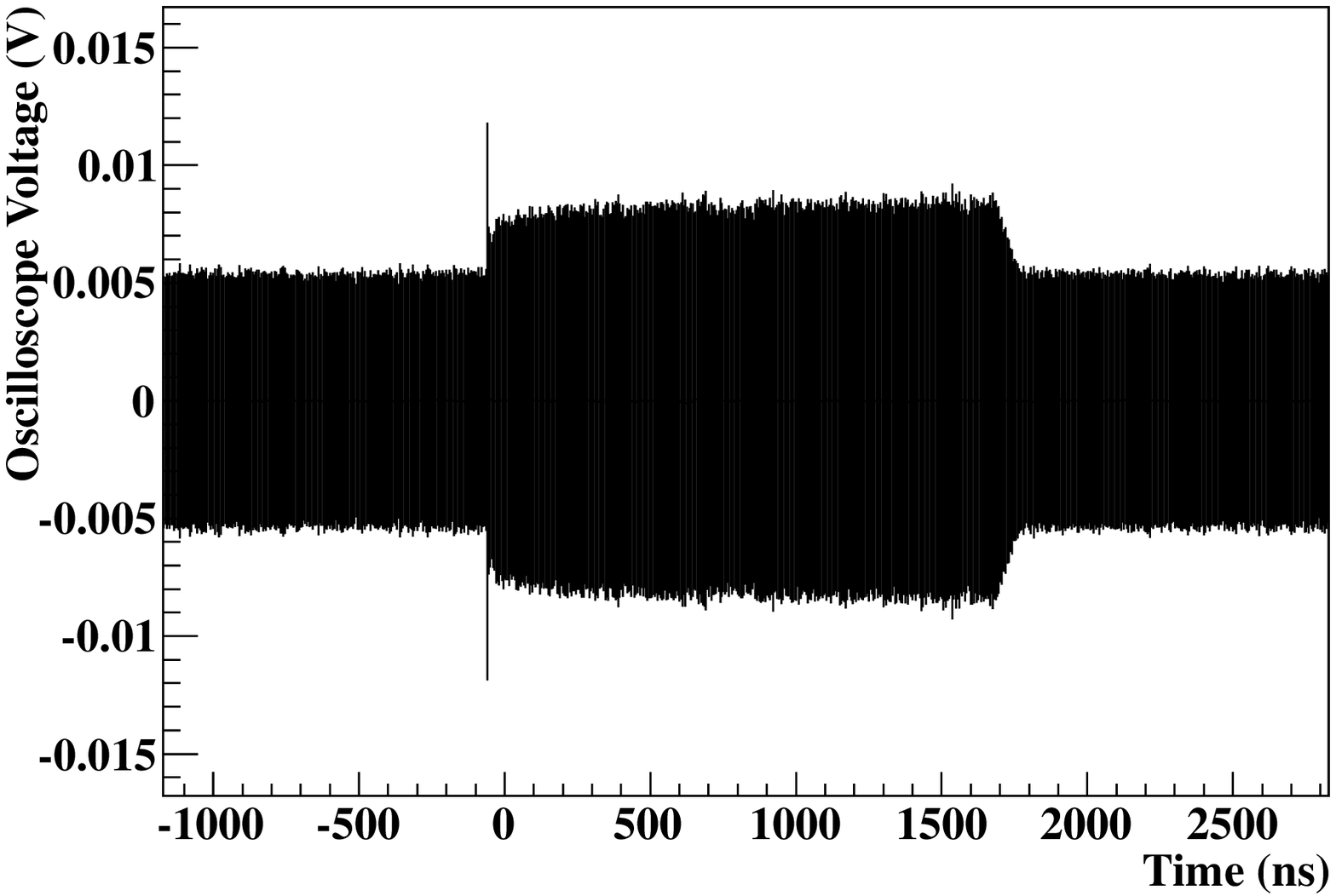}\label{fig3a}
    \hfil
    \includegraphics[height=0.7\columnwidth]{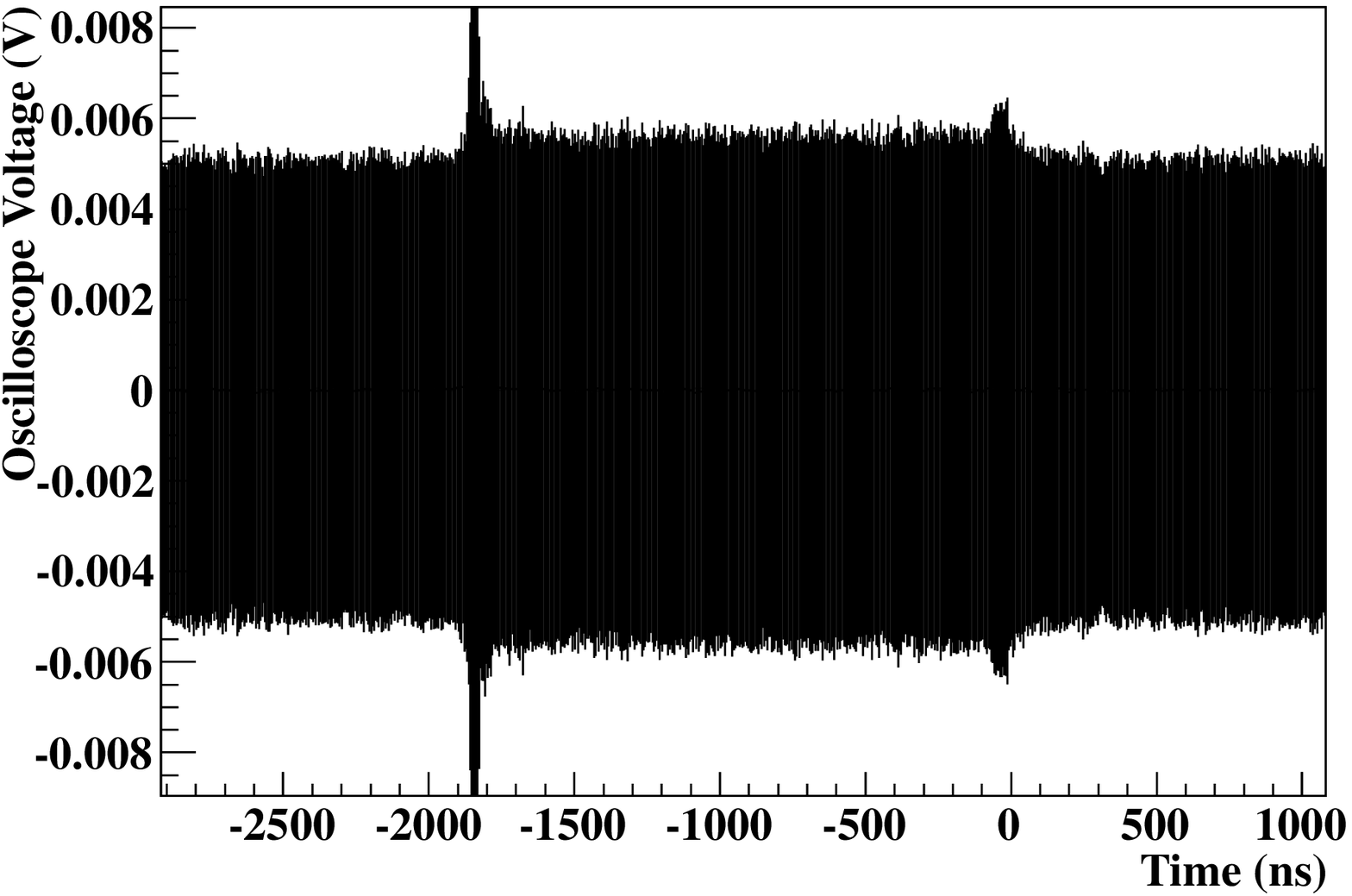} \label{fig3b}
  }
  \caption{Average of 100 traces for the C-band (left) and log-periodic (right) antennas after filtering frequencies of interest with a FFT analysis in a trace by trace basis.
  }
  \label{CI_LowI_trace}
\end{figure*}

A first step in the analysis of the traces is to select only frequencies in the amplified spectral range using a Fast Fourier Transform algorithm. Figure~\ref{CI_LowI_trace} show the average of 100 traces after selecting the frequency range of interest event by event.  For this particular case the beam was pulsed with a rate of 10 Hz and $\sim$2$\mu$s width. In order to determine the power measured in the receivers $P_{\rm measured}$ (the emitted one relies on a frequency calibration of the absolute gain of the whole system that is still pending) the RMS of the recorded voltage over a defined time window $V_{\rm RMS}$ is calculated. Then the distribution of $V_{\rm RMS}$ for the filtered traces in the time range of interest is obtained. The emitted power is finally calculated as:
\begin{equation}
P_{\rm measured}=\left(\frac{<V_{\rm RMS}>^2}{R}\right)_{\rm beam} - \left(\frac{<V_{\rm RMS}>^2}{R}\right)_{\rm bkg}
\end{equation}
\noindent where the first term is calculated over a time period where the beam is active and stable, and the background contribution comes from the first part of the trace. $R$ is the oscilloscope input impedance equal to 50$\Omega$. No radiation was registered in the antennas when the beam was blocked so no additional background subtraction due to emission coming from the accelerator itself is needed.  

Due to the beam structure, the described setup is not suitable to study a possible exponential decay of the signal due to thermalization processes.

\subsection{Scan in beam intensity}
\begin{figure*}[!t]
  \centerline{\includegraphics[width=1.1\columnwidth]{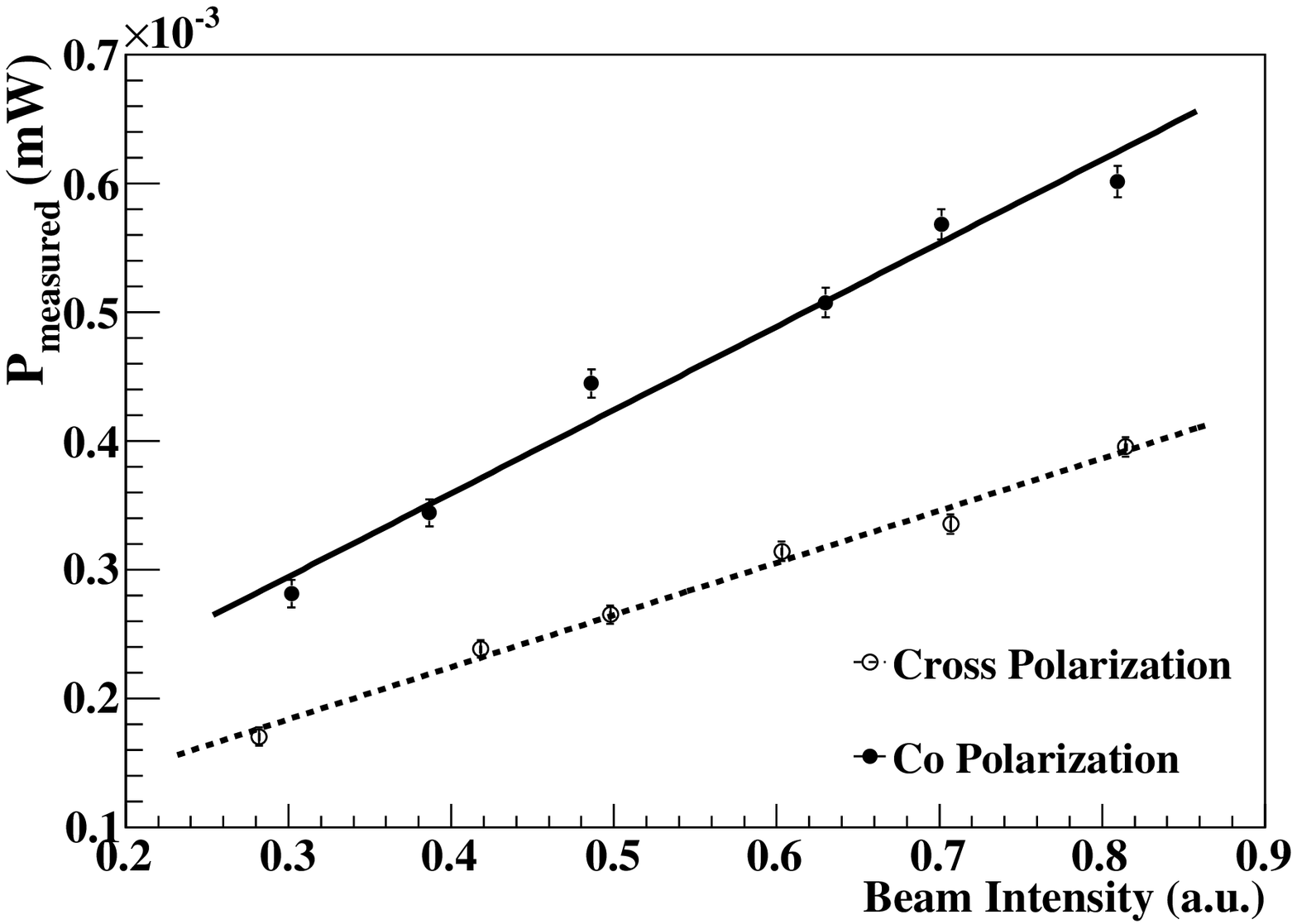}\label{fig4a}
    \hfil
    \includegraphics[width=1.1\columnwidth]{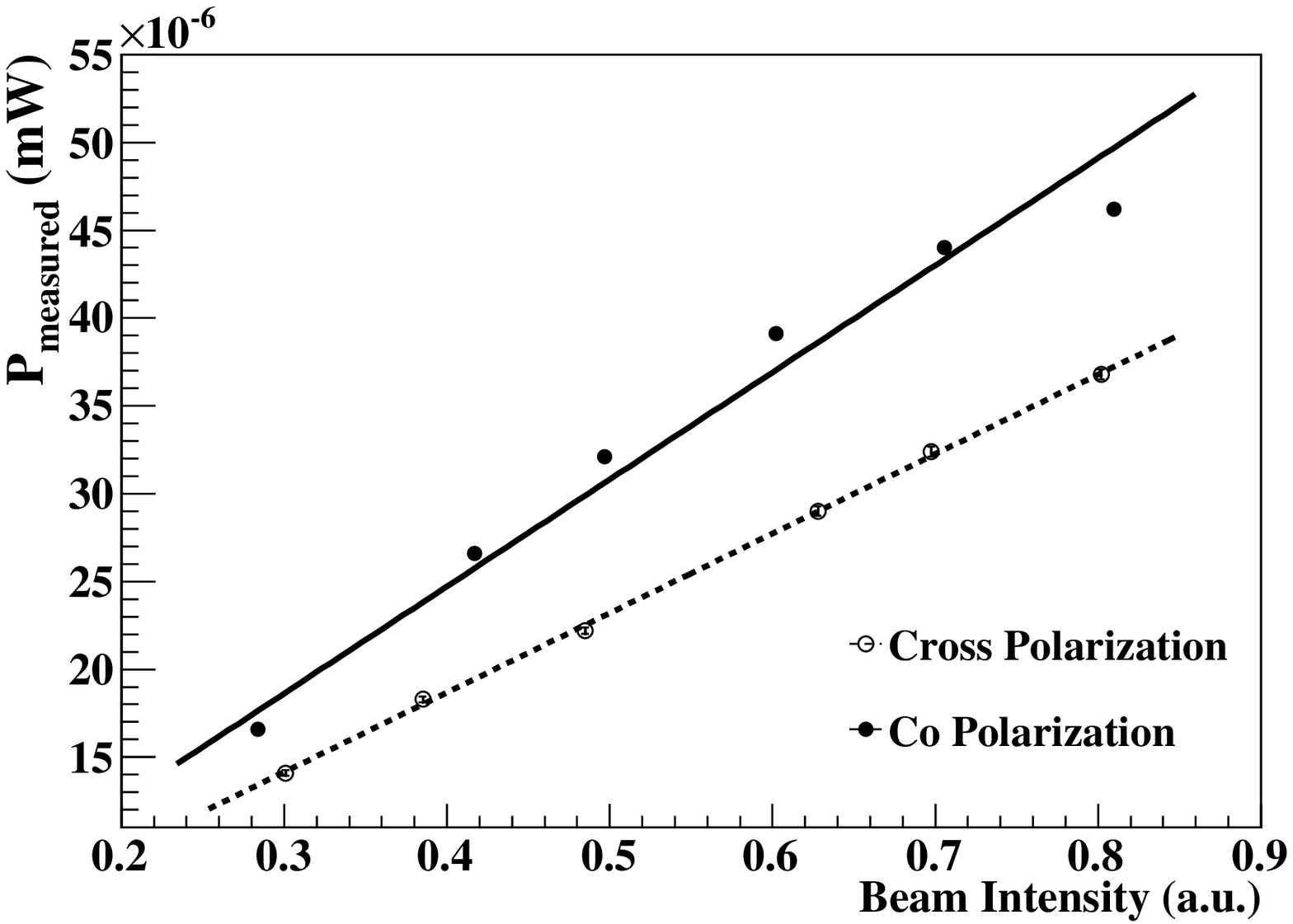} \label{fig4b}
  }
  \caption{Evolution of the measured power $P_{\rm measured}$ as a function of beam energy for C-band (left) and log-periodic (right) antennas.
  }
  \label{linearity}
\end{figure*}

One of the goals of the experiment was to measure the proportionality of the recorded signal with beam energy, i.e. with the number of electrons in the beam. The described analysis was applied to traces measured for $\sim$1$\mu$s width beam pulses with an intensity going from $\sim$5 to 30 $\mu$A\footnote{it corresponds to a beam current between 0.5 and 3 $\mu$C per pulse.}. Results for both C-band and log-periodic antennas located inside the chamber and averaged over 100 traces are shown in figure~\ref{linearity}. Both polarizations, co and cross-polarized with the beam direction, were analyzed. Linear fits to the data are also shown. Preliminary results indicate a linear dependence on beam energy of the power emitted and a difference between polarizations of $\sim30\%$ for C-band and $\sim60\%$ for the low frequency measurements. This difference could be due to the presence of different processes with different degrees of polarization or to near field effects.

To bring some light into the origin of the measured radiation we did another scan in intensity with a thin aluminum foil covering the entry hole in order to block possible contributions from outside radiation. No significant decrease in the total energy deposit inside the chamber due to the additional extra layer is expected since beam electrons will be almost unaltered. We are aware transition radiation can be also generated in this new layer but we expect it to be different to the one generated at the Duraluminum window due to the different geometries.  In the configuration with the hole open, some incidence angles are cut and diffraction effects can take place at the aperture. No significant discrepancies in $P_{\rm measured}$, in the dependence with beam energy or in the difference between polarizations were observed. This result could indicate that the measured signal is indeed produced inside the chamber.

\subsection{Spectrum measurements}

Measurements were also taken in frequency space for both polarizations of C-band and low frequency antennas. Even though final results are still pending on an absolute calibration in frequency to correct for the multiple effects involved in the data acquisition process, a preliminary analysis of the data shows a spectrum compatible with a continuum emission between 1 and 4 GHz. The difference between polarizations observed for the integrated spectral range is still present but a study of this discrepancy as a function of frequency is still ongoing. 


\section{Discussion and future work}

Results presented in this work are preliminary. Final results for the absolute emitted flux, its dependence on frequency as well as the scaling to air showers  will need a frequency calibration of the whole setup and a better understanding of the origin of the measured radiation. For the conditions recreated with a 3 MeV electron beam, the measured signal is consistent with incoherent radiation. The origin of this emission is still unknown due to possible contributions from transition radiation generated at the window placed at exit of the beam.  Another beam test is already scheduled to characterize the emission between 2 and 12 GHz. 





\clearpage

\end{document}